\documentclass[twocolumn,floats,floatfix,aps,pra]{revtex4}
\usepackage{eurosym}
\usepackage{amsfonts}
\usepackage{amssymb}
\usepackage{amsmath}
\usepackage{graphicx}
\usepackage{bm}
\usepackage{color}
\usepackage{epsfig}
\usepackage{ifthen}
\usepackage{subfigure}
\usepackage{physics}
\usepackage[breaklinks,hidelinks, colorlinks=true,linkcolor=blue,citecolor=blue,filecolor=blue,urlcolor=blue]{hyperref}

\begin{document}

\title{Arbitrary polarization retarders and polarization controllers,
constructed from sequences of half-wave and quarter-wave plates}

\author{Hayk L. Gevorgyan}
\affiliation{Center for Quantum Technologies, Faculty of Physics, St. Kliment Ohridski University of Sofia, 5 James Bourchier Blvd., 1164 Sofia, Bulgaria}
\affiliation{Quantum Technologies Division, Alikhanyan National Laboratory (Yerevan Physics Institute), 2 Alikhanyan Brothers St., 0036 Yerevan, Armenia}
\affiliation{Experimental Physics Division, Alikhanyan National Laboratory (Yerevan Physics Institute), 2 Alikhanyan Brothers St., 0036 Yerevan, Armenia}
\author{Andon A. Rangelov}
\affiliation{Center for Quantum Technologies, Faculty of Physics, St. Kliment Ohridski University of Sofia, 5 James Bourchier Blvd., 1164 Sofia, Bulgaria}

\begin{abstract}
We theoretically introduce several types of arbitrary polarization retarders constructed from sequences of half-wave and quarter-wave plates, each rotated at specific angles. By integrating these arbitrary polarization retarders with arbitrary polarization rotators, we develop a versatile device capable of performing arbitrary-to-arbitrary polarization transformations. While some of the proposed devices are documented in the literature, others are novel and, to the best of our knowledge, have not been previously presented. The continuous adjustment of retardance and rotation in these devices is achieved by altering the relative orientation of the wave plates in the sequence.
\end{abstract}

\maketitle

\section{Introduction}


Polarization is one of the essential characteristics of light \cite{Hecht,Wolf,Azzam,Goldstein,Duarte}. The ability to detect and manipulate the polarization state has significant practical value across a range of disciplines. Polarization-sensitive measurement techniques find applications in areas such as stress analysis, ellipsometry, physics, chemistry, biological sciences, astronomy, and more \cite{Pye,Damask,Landolfi}. Additionally, precise control over light polarization plays a critical role in modern display systems and optical communication technologies \cite{Matioli}.

The primary optical tools used to alter polarization are retarders and rotators \cite{Hecht,Wolf,Azzam,Goldstein,Duarte}. Retarders, commonly known as wave plates, function by introducing a defined phase delay—or retardance—between orthogonal polarization components of a light wave. This phase delay enables the transformation of the polarization state, allowing, for example, conversion between linear, circular, and elliptical polarization forms \cite{Hecht,Wolf,Azzam,Goldstein,Duarte}.

Depending on the specific phase shift they provide, retarders are typically classified into two main types: quarter-wave and half-wave plates. A quarter-wave plate introduces a phase delay of one-quarter wavelength, which makes it ideal for interconverting linear and circular polarization. In contrast, a half-wave plate produces a half-wavelength phase shift and is widely used to rotate the orientation of linearly polarized light.

A polarization rotator is an optical element designed to rotate the plane of polarization of linearly polarized light by a constant angle, regardless of the initial orientation. The most widely utilized rotators are Faraday rotators, which use circular birefringence induced by a magnetic field (Faraday effect) \cite{Moeller-book}. Because Faraday rotators are non-reciprocal—that is, they behave differently depending on the direction of light propagation—they are commonly integrated into optical isolators when combined with polarizers and analyzers \cite{Rayleigh1885}. Nonetheless, their drawbacks include large size, high cost, and sensitivity to temperature due to the dispersion of the Verdet constant.

An alternative method for implementing polarization rotation uses naturally optically active crystals like quartz, which exhibit circular birefringence. These devices are commercially available, reciprocal, and offer fixed rotation angles determined by the thickness of the crystal. However, the inability to dynamically tune the rotation angle and strong wavelength dependence of the optical activity limit their flexibility.

Another approach employs twisted nematic liquid crystal cells \cite{Zhuang2000,Chung18}, which are based on the operational principle of LCDs. In sufficiently thick cells, the polarization state of light follows the spatial variation in the orientation of liquid crystal molecules in an adiabatic manner. This results in nearly achromatic behavior, making them suitable for broadband applications. However, the rotation angle is not easily adjustable, and performance can degrade at high optical powers due to thermal effects.

It is also well established that placing two half-wave plates with their optical axes oriented at an angle of $\pi /2+\alpha$ to each other results in polarization rotation by $2\alpha$ \cite{Bhandari,Zhan,Rangelov,Messaadi}. A lesser-known but effective design for a variable polarization rotator consists of a half-wave plate positioned between two quarter-wave plates in the configuration Q($\alpha$)H(0)Q($-\alpha$), where the Q and H denote quarter- and half-wave plates rotated by $\alpha$ and 0 degrees, respectively. This setup acts as a polarization rotator with a total rotation angle of $2\alpha$ \cite{Bhandari}.

In this work, we begin by exploring several configurations of quarter- and half-wave plates that can function as general-purpose retarders. This theoretical framework includes a rotator positioned between a pair of quarter-wave plates with orthogonal fast axes \cite{Ye,Messaadi}. We then build on the well-known principle that any arbitrary polarization transformation can be implemented by combining a retarder and a rotator \cite{Hurvitz}. By integrating generalized retarders and rotators, we design a set of universal polarization controllers. Some of these designs are already described in the literature, while others are, to the best of our knowledge, original contributions presented here for the first time.

\section{Tunable arbitrary polarization rotators}


A polarization rotator that rotates the state of polarization by an angle $\theta$ can be described using the following Jones matrix: 
\begin{equation}
\mathbf{R}(\theta )=\left[ 
\begin{array}{cc}
\cos \theta & \sin \theta \\ 
-\sin \theta & \cos \theta%
\end{array}%
\right] ,
\end{equation}%
Meanwhile, the general Jones matrix for a phase retarder is given by: 
\begin{equation}
\mathbf{J}(\varphi )=\left[ 
\begin{array}{cc}
e^{i\varphi /2} & 0 \\ 
0 & e^{-i\varphi /2}%
\end{array}%
\right] .
\end{equation}%
Here, $\varphi$ denotes the phase difference introduced between orthogonal polarization components. The two most commonly used types of retarders are the quarter-wave plate ($\varphi = \pi /2$) and the half-wave plate ($\varphi = \pi$) \cite{Pye,Damask}. When the polarization axes of the incident light are rotated by an angle $\theta$ relative to the fast and slow axes of the wave plate, the retarder is described by the transformed matrix: 
\begin{equation}
\mathbf{J}_{\theta }(\varphi )=\mathbf{R}(-\theta )\mathbf{J}(\varphi )%
\mathbf{R}(\theta ).  \label{retarder}
\end{equation}

A general polarization rotator can be realized using two half-wave plates \cite{Bhandari,Zhan,Rangelov,Messaadi}, where the optical axes are offset by a relative angle of $\pi/2 + \alpha/2$. The resulting transformation is: 
\begin{equation}
\mathbf{J}_{\alpha /2+\pi /2}(\pi )\mathbf{J}_{0}(\pi )=\left[ 
\begin{array}{cc}
\cos \alpha & \sin \alpha \\ 
-\sin \alpha & \cos \alpha%
\end{array}%
\right] .  \label{rotator1}
\end{equation}%
This outcome follows from the mathematical property that the composition of two reflection matrices—here represented by Jones matrices of half-wave plates—yields a rotation matrix. An alternative method for constructing an arbitrary rotator involves a half-wave plate sandwiched between two quarter-wave plates, arranged as follows \cite{Bhandari}:

\begin{equation}
\mathbf{J}_{\alpha /2}(\pi /2)\mathbf{J}_{\pi /2}(\pi )\mathbf{J}_{-\alpha
/2}(\pi /2)=\left[ 
\begin{array}{cc}
\cos \alpha & \sin \alpha \\ 
-\sin \alpha & \cos \alpha%
\end{array}%
\right] .  \label{rotator2}
\end{equation}%
Both configurations described above serve as arbitrary polarization rotators. The first configuration in Eq.~(\ref{rotator1}) requires adjusting only a single wave plate to tune the rotation angle, while the second configuration in Eq.~(\ref{rotator2}) involves the rotation of two wave plates to achieve the same effect.

\section{Tunable arbitrary polarization retarders}


In this section, we outline the design of tunable polarization retarders using the arbitrary polarization rotators introduced earlier (see Eq.~(\ref{rotator1}) and Eq.~(\ref{rotator2})). As previously discussed in Refs.~\cite{Ye, Messaadi}, inserting a rotator between two quarter-wave plates oriented at $90^\circ$ to each other produces a general retarder, where the retardance is twice the rotation angle of the embedded rotator. The rotator itself can be constructed either from a pair of half-wave plates (as in Eq.~(\ref{rotator1})) or from a half-wave plate sandwiched between two quarter-wave plates (as in Eq.~(\ref{rotator2})). Based on these principles, we propose the following two implementations for arbitrary polarization retarders:
\begin{widetext}
\begin{subequations}
\label{arbitrary retarders}
\begin{eqnarray}
\mathbf{J}_{0}(\alpha ) &=&\mathbf{J}_{\frac{\pi }{4}}(\pi /2)\mathbf{J}%
_{\alpha /4+\pi /2}(\pi )\mathbf{J}_{0}(\pi )\mathbf{J}_{-\frac{\pi }{4}%
}(\pi /2),  \label{Messaadi case} \\
\mathbf{J}_{0}(\alpha ) &=&\mathbf{J}_{\frac{\pi }{4}}(\pi /2)\mathbf{J}%
_{\alpha /4}(\pi /2)\mathbf{J}_{\pi /2}(\pi )\mathbf{J}_{-\alpha /4}(\pi /2)%
\mathbf{J}_{-\frac{\pi }{4}}(\pi /2).  \label{Bhandari case}
\end{eqnarray}
\end{subequations}
\end{widetext}

Equation~(\ref{Messaadi case}) was recently employed by Messaadi et al.~\cite{Messaadi} to construct a tunable and broadband linear polarization retarder. In contrast, Eq.~(\ref{Bhandari case}) appears to be unreported in existing literature, though it includes additional wave plates relative to Eq.~(\ref{Messaadi case}). Next, we show that both expressions in Eqs.~(\ref{arbitrary retarders}) can be simplified further.

Since the result of a polarization rotator depends only on the relative orientation between the wave plates, the angles $\theta_1$ and $\theta_2$ in Eqs.~(\ref{arbitrary retarders}) can be absorbed into the system. Consequently, the retarders can be rewritten as:
\begin{widetext}
\begin{subequations}
\label{arbitrary retarders 2}
\begin{eqnarray}
\mathbf{J}_{0}(\alpha ) &=&\mathbf{J}_{\frac{\pi }{4}}(\pi /2)\mathbf{J}%
_{\alpha /4+\pi /2+\theta _{1}}(\pi )\mathbf{J}_{\theta _{1}}(\pi )\mathbf{J}%
_{-\frac{\pi }{4}}(\pi /2),  \label{Messaadi case 2} \\
\mathbf{J}_{0}(\alpha ) &=&\mathbf{J}_{\frac{\pi }{4}}(\pi /2)\mathbf{J}%
_{\alpha /4+\theta _{2}}(\pi /2)\mathbf{J}_{\pi /2+\theta _{2}}(\pi )\mathbf{%
J}_{-\alpha /4+\theta _{2}}(\pi /2)\mathbf{J}_{-\frac{\pi }{4}}(\pi /2).
\label{Bhandari case 2}
\end{eqnarray}
\end{subequations}
\end{widetext}

By selecting specific values for the free parameters, namely $\theta_1 = \pi/4$ and $\theta_2 = \pi/4 + \alpha/4$, and applying the identities $\mathbf{J}_{\frac{\pi }{4}}(\pi )\mathbf{J}_{-\frac{\pi }{4}}(\pi /2)=\mathbf{J}_{\frac{\pi }{4}}(\pi /2)$ and $\mathbf{J}_{\frac{\pi }{4}}(\pi /2)\mathbf{J}_{-\frac{\pi }{4}}(\pi /2)=\hat{1}$, we arrive at simplified expressions for the arbitrary retarders:

\begin{subequations}
\label{final arbitrary retarders}
\begin{eqnarray}
\mathbf{J}_{0}(\alpha ) &=&\mathbf{J}_{\frac{\pi }{4}}(\pi /2)\mathbf{J}%
_{\alpha /4+3\pi /4}(\pi )\mathbf{J}_{\frac{\pi }{4}}(\pi /2),
\label{Evans case} \\
\mathbf{J}_{0}(\alpha ) &=&\mathbf{J}_{\frac{\pi }{4}}(\pi /2)\mathbf{J}%
_{\alpha /2+\pi /4}(\pi /2)\mathbf{J}_{3\pi /4+\alpha /4}(\pi ).
\label{unknown case}
\end{eqnarray}
\end{subequations}

The retarder described by Eq.~(\ref{Evans case}) is known in the literature as Evans’s retarder or phase shifter \cite{Evans49, Damask}, while the structure in Eq.~(\ref{unknown case}) has not, to our knowledge, been previously reported. It is worth emphasizing that the final configurations in Eqs.~(\ref{final arbitrary retarders}) utilize fewer optical components compared to those in Eqs.~(\ref{arbitrary retarders}), making them more practical and efficient for experimental implementation.

\section{Tunable arbitrary polarization controllers}


To realize arbitrary polarization controllers, we rely on the principle that any polarization transformation can be decomposed into a combination of a retarder and a rotator \cite{Hurvitz}. This concept translates into multiplying the Jones matrices of arbitrary rotators (given in Eqs.~(\ref{rotator1}) and (\ref{rotator2})) with those of the tunable retarders (from Eqs.~(\ref{final arbitrary retarders})). Since matrix multiplication is not commutative, the sequence in which the elements are applied affects the outcome, leading to two distinct possibilities:

\begin{eqnarray}
\mathbf{J}^{R}(\beta ,\alpha ) &=&\mathbf{R}(\beta )\mathbf{J}_{0}(\alpha ), \\
\mathbf{J}^{L}(\beta ,\alpha ) &=&\mathbf{J}_{0}(\alpha )\mathbf{R}(\beta ).
\end{eqnarray}

Here, $\mathbf{J}^{R}(\beta ,\alpha )$ represents a polarization controller where the retarder is applied first, followed by the rotator, while $\mathbf{J}^{L}(\beta ,\alpha )$ corresponds to the reverse order—first the rotator, then the retarder. The rotator $\mathbf{R}(\beta)$ can be constructed using either of the two approaches given in Eqs.~(\ref{rotator1}) and (\ref{rotator2}), and $\mathbf{J}_{0}(\alpha )$ is an arbitrary retarder as given in Eqs.~(\ref{final arbitrary retarders}). This results in the following eight configurations for polarization controllers:

\begin{widetext}
\begin{subequations}
\label{polarization controllers}
\begin{eqnarray}
\mathbf{J}^{R}(\beta ,\alpha ) &=&\mathbf{J}_{\beta /2+\pi /2+\delta _{1}}(\pi )\mathbf{J}_{\delta _{1}}(\pi )\mathbf{J}_{\frac{\pi }{4}}(\pi /2)\mathbf{J}_{\alpha /4+3\pi /4}(\pi )\mathbf{J}_{\frac{\pi }{4}}(\pi /2), \\
\mathbf{J}^{R}(\beta ,\alpha ) &=&\mathbf{J}_{\beta /2+\pi /2+\delta _{2}}(\pi )\mathbf{J}_{\delta _{2}}(\pi )\mathbf{J}_{\frac{\pi }{4}}(\pi /2)\mathbf{J}_{\alpha /2+\pi /4}(\pi /2)\mathbf{J}_{3\pi /4+\alpha /4}(\pi ), \\
\mathbf{J}^{R}(\beta ,\alpha ) &=&\mathbf{J}_{\beta /2+\delta _{3}}(\pi /2)\mathbf{J}_{\pi /2+\delta _{3}}(\pi )\mathbf{J}_{-\beta /2+\delta _{3}}(\pi /2)\mathbf{J}_{\frac{\pi }{4}}(\pi /2)\mathbf{J}_{\alpha /4+3\pi /4}(\pi )\mathbf{J}_{\frac{\pi }{4}}(\pi /2), \\
\mathbf{J}^{R}(\beta ,\alpha ) &=&\mathbf{J}_{\beta /2+\delta _{4}}(\pi /2)\mathbf{J}_{\pi /2+\delta _{4}}(\pi )\mathbf{J}_{-\beta /2+\delta _{4}}(\pi /2)\mathbf{J}_{\frac{\pi }{4}}(\pi /2)\mathbf{J}_{\alpha /2+\pi /4}(\pi /2)\mathbf{J}_{3\pi /4+\alpha /4}(\pi ), \\
\mathbf{J}^{L}(\beta ,\alpha ) &=&\mathbf{J}_{\frac{\pi }{4}}(\pi /2)\mathbf{J}_{\alpha /4+3\pi /4}(\pi )\mathbf{J}_{\frac{\pi }{4}}(\pi /2)\mathbf{J}_{\beta /2+\pi /2+\delta _{5}}(\pi )\mathbf{J}_{\delta _{5}}(\pi ), \\
\mathbf{J}^{L}(\beta ,\alpha ) &=&\mathbf{J}_{\frac{\pi }{4}}(\pi /2)\mathbf{J}_{\alpha /2+\pi /4}(\pi /2)\mathbf{J}_{3\pi /4+\alpha /4}(\pi )\mathbf{J}_{\beta /2+\pi /2+\delta _{6}}(\pi )\mathbf{J}_{\delta _{6}}(\pi ), \\
\mathbf{J}^{L}(\beta ,\alpha ) &=&\mathbf{J}_{\frac{\pi }{4}}(\pi /2)\mathbf{J}_{\alpha /4+3\pi /4}(\pi )\mathbf{J}_{\frac{\pi }{4}}(\pi /2)\mathbf{J}_{\beta /2+\delta _{7}}(\pi /2)\mathbf{J}_{\pi /2+\delta _{7}}(\pi )\mathbf{J}_{-\beta /2+\delta _{7}}(\pi /2), \\
\mathbf{J}^{L}(\beta ,\alpha ) &=&\mathbf{J}_{\frac{\pi }{4}}(\pi /2)\mathbf{J}_{\alpha /2+\pi /4}(\pi /2)\mathbf{J}_{3\pi /4+\alpha /4}(\pi )\mathbf{J}_{\beta /2+\delta _{8}}(\pi /2)\mathbf{J}_{\pi /2+\delta _{8}}(\pi )\mathbf{J}_{-\beta /2+\delta _{8}}(\pi /2),
\end{eqnarray}
\end{subequations}
\end{widetext}

These expressions reflect equivalent transformations for different values of $\delta_i$, due to the symmetry of rotators under angular shifts. By assigning the free parameters as follows:  
$\delta _{1}=\delta _{2}=-\pi /4$,  
$\delta _{3}=\delta _{4}=\beta /2-\pi /4$,  
$\delta _{5}=-3\pi /4-\beta /2$,  
$\delta _{6}=3\pi /4+\alpha /4-\beta /2$,  
$\delta _{7}=-\pi /4-\beta /2$,  
$\delta _{8}=\pi /4+\alpha /4-\beta /2$,  
we can simplify the above expressions to:

\begin{widetext}
\begin{subequations}
\label{polarization controllers 3}
\begin{eqnarray}
\mathbf{J}^{R}(\beta ,\alpha ) &=&\mathbf{J}_{\beta /2+\pi /4}(\pi )\mathbf{J}_{-\frac{\pi }{4}}(\pi /2)\mathbf{J}_{\alpha /4+3\pi /4}(\pi )\mathbf{J}_{\frac{\pi }{4}}(\pi /2), \label{Simon and Mukunda} \\
\mathbf{J}^{R}(\beta ,\alpha ) &=&\mathbf{J}_{\beta /2+\pi /4}(\pi )\mathbf{J}_{-\frac{\pi }{4}}(\pi /2)\mathbf{J}_{\alpha /2+\pi /4}(\pi /2)\mathbf{J}_{3\pi /4+\alpha /4}(\pi ), \label{Case A} \\
\mathbf{J}^{R}(\beta ,\alpha ) &=&\mathbf{J}_{\beta -\pi /4}(\pi /2)\mathbf{J}_{\pi /4+\beta /2}(\pi )\mathbf{J}_{\alpha /4+3\pi /4}(\pi )\mathbf{J}_{\frac{\pi }{4}}(\pi /2), \label{Case B} \\
\mathbf{J}^{L}(\beta ,\alpha ) &=&\mathbf{J}_{\frac{\pi }{4}}(\pi /2)\mathbf{J}_{\alpha /2+\pi /4}(\pi /2)\mathbf{J}_{3\pi /4+\alpha /4-\beta /2}(\pi ), \label{3 matrixes} \\
\mathbf{J}^{L}(\beta ,\alpha ) &=&\mathbf{J}_{\frac{\pi }{4}}(\pi /2)\mathbf{J}_{\alpha /2+\pi /4}(\pi /2)\mathbf{J}_{3\pi /4+\alpha /4}(\pi /2)\mathbf{J}_{3\pi /4+\alpha /4-\beta /2}(\pi )\mathbf{J}_{\pi /4-\beta +\alpha /4}(\pi /2). \label{5 matrixes}
\end{eqnarray}
\end{subequations}
\end{widetext}

Among these, some cases are identical. Removing duplicates gives us a concise list of unique polarization controller configurations.

If the input state is purely circular (left or right), applying a rotator alone has no effect, as circular polarization remains unchanged under rotation. Hence, to achieve complete control over any polarization transformation, it is necessary to first apply a retarder. In such cases, the $\mathbf{J}^{L}(\beta ,\alpha )$ schemes must be inverted.

Notably, the setup described by Eq.~(\ref{5 matrixes}) involves five wave plates, making it less efficient. On the other hand, Eq.~(\ref{3 matrixes}) only requires three elements, making it the simplest and most practical option; this design was previously proposed by Simon and Mukunda \cite{Mukunda}. Eq.~(\ref{Simon and Mukunda}) was also suggested in earlier work by the same authors \cite{Simon}. However, to the best of our knowledge, configurations described in Eq.~(\ref{Case A}) and Eq.~(\ref{Case B}) are novel and have not been reported previously, representing new contributions to the design of tunable polarization controllers.


\section{Summary}


To summarize, we have presented a theoretical framework for designing various forms of arbitrary polarization retarders using specific arrangements of half-wave and quarter-wave plates set at carefully chosen orientation angles. By integrating these configurable retarders with generalized polarization rotators, we have constructed flexible optical systems capable of implementing transformations between any two polarization states. Although certain configurations align with previously known setups from the literature, others appear to be new contributions not previously reported. These devices allow for continuous tuning of both retardation and rotation angles by simply modifying the angular relationships between the constituent wave plates.


\section*{Acknowledgments}

This research is partially supported by the Bulgarian national plan for
recovery and resilience, contract BG-RRP-2.004-0008-C01 SUMMIT: Sofia
University Marking Momentum for Innovation and Technological Transfer,
project number 3.1.4.

\end{document}